Editorial for the Proceedings of the Workshop
# Knowledge Maps and Information Retrieval (KMIR2014)
at Digital Libraries 2014

Peter Mutschke*, Philipp Mayr*, Andrea Scharnhorst**

*GESIS – Leibniz Institute for the Social Sciences, Cologne, Germany
** DANS, Royal Netherlands Academy of Arts and Sciences, Amsterdam, The Netherlands
`peter.mutschke@gesis.org`

**Abstract.** Knowledge maps are promising tools for visualizing the structure of large-scale information spaces, but still far away from being applicable for searching. The first international workshop on "Knowledge Maps and Information Retrieval (KMIR)", held as part of the International Conference on Digital Libraries 2014 in London, aimed at bringing together experts in Information Retrieval (IR) and knowledge mapping in order to discuss the potential of interactive knowledge maps for information seeking purposes.

## 1 Introduction

The success of an information system depends mainly on its ability to properly support interaction between users and information. Current information systems, however, show as a particular point of failure the vagueness between user search terms and the knowledge orders of the information space in question [1,2]. Studies in interactive information seeking behavior have confirmed that the ability to browse an information space and observe similarities and dissimilarities between information objects is crucial for accidental encountering and the creative use of information [3,4]. This is in particular true for heterogeneous information spaces. Some kind of guided searching therefore becomes more and more important in order to precisely discover information without knowing the right search terms. Yet, this seems to remain the weakest point of interactive information systems [5-7].

Knowledge mapping encompasses all attempts to use visualizations to gain insights into the structure and evolution of large-scale information spaces. Knowledge maps can take very different forms of visualizing the structure of information spaces [8-14]. As an activity performed in very different disciplines – and often independently from each other – it stands in line with the dominance of the visual in our culture [15]. Knowledge maps of Digital Library (DL) collections are promising navigation tools through knowledge spaces but – to the best of our knowledge – still far away from being applicable as search interfaces for DLs. Most maps are made for special purposes, are static, and usually not interactive [16].

In interactive information systems the use of visual elements to enhance information seeking and discovery is a recurring research issue. However, not much of the experiences made in knowledge mapping have ever been implemented in online interfaces to DL collections [17], nor is there a stable and continuous knowledge exchange between the "map makers" on the one hand and the Information Retrieval (IR) specialists on the other hand. Thus, there is also a lack of models that properly combine insights of the two strands, which are driven by quite different epistemic perspectives.

This first international workshop on "Knowledge Maps and Information Retrieval (KMIR)"[1] aimed at bringing together these two communities: experts in IR reflecting on visually enhanced search interfaces and experts in knowledge mapping reflecting on visualizations of the content of a collection that might also present – visually – a context for a search term. The intention of the workshop is to raise awareness of the potential of interactive knowledge maps for information seeking purposes and to create a common ground for experiments aiming at the incorporation of knowledge maps into IR models at the level of the user interface. The major focus of the workshop was on the question of how knowledge maps can be utilized for scholarly information seeking in large information spaces. This issue is closely related to the COST action "Analyzing the dynamics of information and knowledge landscapes" (KNOWeSCAPE)[2] which aims at implementing new navigation and search strategies based on insights of the complex nature of knowledge spaces as well as visualization principles for knowledge maps.

The long-term research goal is to develop and evaluate new approaches for combining knowledge mapping and IR. More specifically, we address questions such as:

- What are appropriate interactive knowledge maps for IR systems?
- How can knowledge maps be utilized for information seeking purposes?
- How to locate an information need on a knowledge map?
- How can interaction with knowledge maps be transformed into IR tasks?
- Can knowledge maps improve searching in large-scale information spaces?
- And the other way around: Can insights from IR also improve knowledge mapping itself?

The half-day KMIR workshop was held in conjunction with the COST Action KNOWeSCAPE and as part of the International Conference on Digital Librar-ies 2014[3] - ACM/IEEE Joint Conference on Digital Libraries (JCDL 2014) and International Conference on Theory and Practice of Digital Libraries (TPDL 2014) in London, 11th September 2014.

---
[1] http://www.gesis.org/en/events/events-archive/conferences/kmir2014
[2] http://knowescape.org/

## 2      Overview of papers

The workshop gave floor to one keynote by **André Skupin** (San Diego State University, USA) and ten presentations spanning a wide range of questions around knowledge maps to be implemented as search interfaces for Digital Libraries. Skupin's keynote on "Managing Domain Knowledge: Ontology, Visualization, and Beyond" focussed on high-dimensional models of so-called Bodies of Knowledge (BoKs) that can be leveraged within interactive visualization. What followed was a search for best ways to create knowledge maps from different kind of data by different kind of visualization methods and metaphors (such as graph visualization of relational data, radial histograms of multi-faceted data, or visualization of knowledge structures by geographic metaphors) as well as perspectives to use knowledge maps as search interfaces. One thread through the presentations was to provide the user with an overview. Another shared topic concerned issues of design, such as fonts, colors, or use of the space.

The paper "Dewey Decimal Classification Based Concept Visualization for Information Retrieval" by **Ahn, Lin & Khoo** introduces a novel visual search interface that dynamically exploits Dewey Decimal Classification annotation for visualizing the knowledge structure of search results. The interface provides interactive manipulation, exploration and filtering of concepts and links at different levels.

The paper "Creating knowledge maps using Memory Islands" by **Yang & Ganascia** describes the idea of Memory Islands which are spatial cartographic representations of a given hierarchical knowledge structure (e.g., an ontology). With the help of interactive functions (e.g. pan, zoom, search and filter) the user can navigate through the artificial landscape.

The paper "Using Font Attributes in Knowledge Maps and Information Retrieval" by **Brath & Banissi** demonstrates the usefulness of simple font attributes for facilitating text skimming and refinement. The paper shows that font-attribute-focused visualization techniques can increase data density for information gathering, fact finding and other lookup strategies.

The paper "Augmenting Citation Chain Aggregation with Article Maps" by **Cribbin** presents an experimental system providing an interactive article map based on the citation network of "pearls" of known articles. In this map arti-cles are arranged according to their content similarity. The paper describes a scenario showing how the map can be used for searching.

---
[3] http://www.dl2014.org/

The paper "Creating a Knowledge Map for the Research Lifecycle" by **Deng & Hu** presents a mind map like visualization of a research lifecycle. The map provides interactive nodes which lead to different resources on the Web.

The paper "How can heat maps of indexing vocabularies be utilized for information seeking purposes?" by **Mutschke & Haddou ou Moussa** discusses a heat map like visualization of a term co-occurrence matrix which can be used as a visual navigation tool.

The paper "Towards a Visualization of Multi-faceted Search Results" by **Alsallakh, Miksch & Rauber** proposes a visualization approach for multi-faceted data based on radial sets to support the user in multi-faceted search.

The paper "Introducing a User Interface with an Entity-Strategy-based Approach for Exploring Document Collections" by **Hienert & van Hoek** discusses a search path graph approach that provides powerful interactive sub-units such as encapsulated search strategies.

The paper "VISFACET: Facet Visualization Module for Modern Library Catalogues" by **Allalouf, Mendelsson & Mishustin** introduces a visual search extension for the library system VUFind providing a network visualization of different kinds of facets.

The paper "Using Extended Abstract Tasks for Evaluating Visual User-Interfaces" by **Triebel, Klas & Hemmje** discusses methods for evaluating visual user interfaces and proposes an evaluation model featuring information visualization aspects, such as visual tasks.

## 3    Outlook

The fruitful discussion at the end of the workshop turned out that for pushing the implementation of knowledge maps in DLs testbeds, user studies and evaluations are strongly needed. A bridge between IR, information studies and information visualization would be a first step to activate a community to join forces for the implementation of knowledge maps in DLs. Comparing and evaluating such interface pilots is not easy. The availability of new IR test collections that contain citation and bibliographic information like the iSearch collection [18] or the ACL collection [19] provides an interesting playground for developing and evaluating combined models of IR and knowledge mapping for scholarly searching.

Apart from studying particular implementations of knowledge maps as search interfaces there is also a growing need to create interfaces which embody the concept of a macroscope. The term macroscope was coined by Katy Börner. She writes: "Macroscopes provide a 'vision of the whole', helping us

'synthesize' the related elements and detect patterns, trends, and outliers while granting access to myriad details. Rather than making things larger or smaller, macroscopes let us observe what is at once too great, slow, or complex for the human eye and mind to notice and comprehend." [20]. Thus, a major challenge is to be seen in the development and evaluation of visual means providing an overview of where we are, where we came from, and where we might go.

**Acknowledgement**: Part of this work has been funded by the COST Action TD1210 KNOWeSCAPE.